\documentclass{article}

 \usepackage[dblblindworkshop,final]{neurips_2025}

\workshoptitle{AI for Music}

\usepackage{graphicx}
\usepackage{booktabs}
\usepackage{multirow}
\usepackage{multicol}
\usepackage{amsmath}
\usepackage{wrapfig}
\usepackage[utf8]{inputenc} 
\usepackage[T1]{fontenc}    
\usepackage{hyperref}       
\usepackage{url}            
\usepackage{booktabs}       
\usepackage{amsfonts}       
\usepackage{nicefrac}       
\usepackage{microtype}      
\usepackage{xcolor}         
\usepackage{wrapfig}
\title{Adapting Speech Language Model to \\ Singing Voice Synthesis}

\author{%
  \textbf{Yiwen Zhao}$^{1}$ \quad \textbf{Jiatong Shi}$^{1}$ \quad \textbf{Jinchuan Tian}$^{1}$ \quad \textbf{Yuxun Tang}$^{2}$ \quad \\ \textbf{Jiarui Hai}$^{3}$ \quad \textbf{Jionghao Han}$^{1}$ \quad \textbf{Shinji Watanabe}$^{1}$ \\
  $^{1}$ Carnegie Mellon University \quad $^{2}$ Renmin University of China \quad $^{3}$ John Hopkins University
}

\begin{document}

\maketitle

\vspace{-5mm}
\begin{abstract}

Speech Language Models (SLMs) have recently emerged as a unified paradigm for addressing a wide range of speech-related tasks, including text-to-speech (TTS), speech enhancement (SE), and automatic speech recognition (ASR). However, the generalization capability of large-scale pre-trained SLMs remains underexplored. In this work, we adapt a 1.7B parameter TTS pretrained SLM for singing voice synthesis (SVS), using only a 135-hour synthetic singing corpus, ACE-Opencpop. Building upon the ESPNet-SpeechLM, our recipe involves the following procedure: (1) tokenization of music score conditions and singing waveforms, (2) multi-stream language model token prediction, (3) conditional flow matching-based mel-spectrogram generation. (4) a mel-to-wave vocoder. Experimental results demonstrate that our adapted SLM generalizes well to SVS and achieves performance comparable to leading discrete token-based SVS models. Project page with sample and code is available\footnote{\label{fn:projectpage}https://tsukasane.github.io/SLMSVS/}.
\end{abstract}
\vspace{-5mm}

\section{Introduction}
\label{sec:intro}

\vspace{-2mm}

Large language models (LLMs) have attracted considerable attention in recent years due to their ability to unify representations across diverse data modalities. This unifying capability enables a single model architecture to scale effectively and generalize across a broad range of tasks. By adopting consistent paradigms for data processing and prediction, LLMs can be efficiently adapted to downstream applications, even in low-resource scenarios.

In the speech domain, prior research has primarily pursued two approaches: (1) fine-tuning language models pretrained on text for speech-related tasks, or (2) training language models directly on speech data. The latter approach, often referred to as Speech Language Models (SLMs), tends to capture fine-grained acoustic characteristics more effectively due to its native exposure to audio signals. However, SLMs are inherently data-intensive, requiring large-scale paired datasets. Consequently, most existing pre-training efforts focus on well-resourced tasks such as text-to-speech (TTS) and automatic speech recognition (ASR).

In contrast, singing voice synthesis (SVS) presents additional challenges. The input consists of richly structured musical scores, including phoneme-level lyrics, precise duration annotations, and MIDI notes. The output is vocal singing that must be both musically and phonetically faithful to these conditions. Compared to TTS, publicly available SVS datasets are far more limited due to restrictive licensing and the labor-intensive nature of score annotation.

\begin{figure*}[t]
\centering
\vspace{-14mm}
\includegraphics[width=13.8cm]{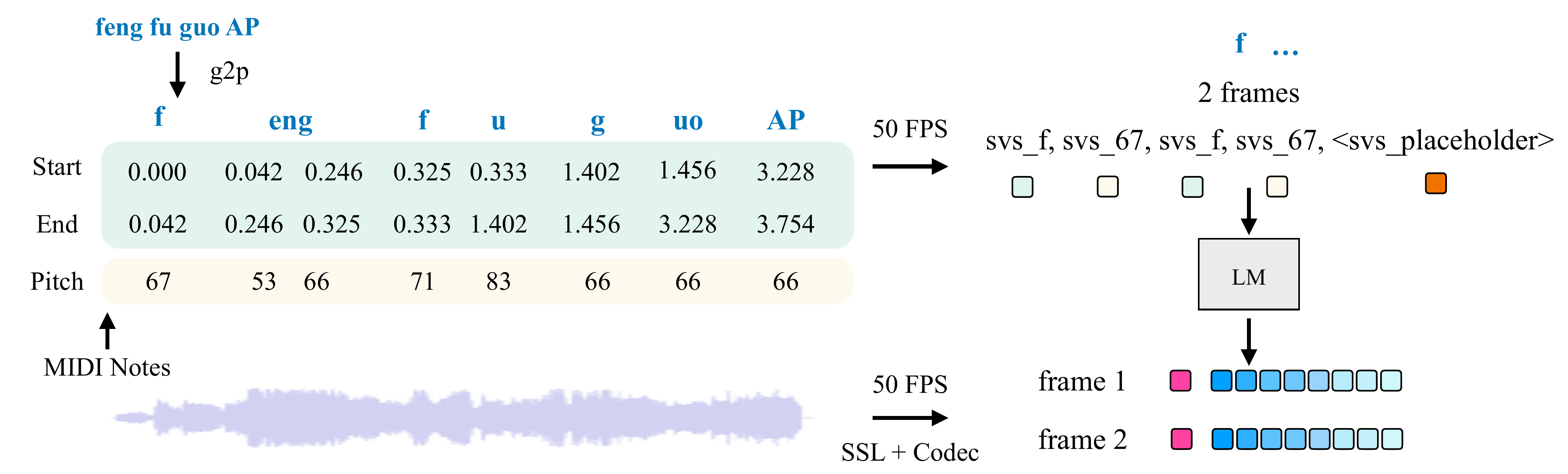}
\caption{\small Illustration of music notes tokenization and waveform tokenization. The phoneme, duration, and MIDI are quantized to 50FPS discrete tokens, appended to the TTS vocabulary. The audio tokens are obtained by a pretrained codec encoder and SSL model, with each frame represented by a concatenation of one SSL token and eight codec tokens.}
\vspace{-2mm}
\label{fig:tokenization}
\vspace{-4mm}
\end{figure*}

To explore the generalization capability of SLMs, we propose adapting a TTS-pretrained SLM to the SVS task. We first tokenize the input music score and target singing waveforms as shown in Fig.~\ref{fig:tokenization}, formulating a multi-stream token prediction task to fine-tune the LM. The predicted tokens, including SSL and multi-layer codec tokens, are able to be separated and decoded to a waveform using the pretrained codec decoder. However, our primary experiment shows that the raw predicted tokens are noisy, as shown in Tab.~\ref{tab_abla}, with resulting waveforms exhibiting temporal discontinuities, particularly at token boundaries, leading to perceptual glitches and unnatural transitions~\textsuperscript{\ref{fn:projectpage}}. Moreover, as the codec model~\cite{espnet-codec} is pretrained on speech data, it lacks the ability to faithfully resynthesize singing, resulting in a performance upper bound set by the decoder side.

To alleviate the unsatisfactory performance introduced by the codec decoder and the noisy tokens, we use a conditional flow matching model, converting the source Gaussian noise to the target mel spectrogram conditioned on the codec, and additionally train a vocoder~\cite{hifigan} that is consistent with the codec STFT parameters. This optimization enables high-quality singing synthesis while addressing the limitations posed by data scarcity. Considering the expressiveness of synthesized singing, we strengthen the condition of pitch information again through the flow matching process, which improves the melodious fidelity. 

Empirical results demonstrate that this second-stage refinement improves synthesis quality, yielding smoother transitions and enhanced pitch accuracy. Overall, our framework enables SLM-based SVS to achieve performance comparable to leading discrete SVS systems. A detailed introduction of related works is attached in appendix~\ref{sec_rw}.

\vspace{-4mm}

\section{Methodology}
\vspace{-2mm}
Given the music score $M:= (M^{\text{ph}}, M^{\text{pi}}, M^{\text{du}})$, SVS targets to generate a human singing phrase $Y \in \mathbb{R}^{T}$ that align with $M$, where $T$ is the number of samples in the waveform, $M^{\text{ph}}\in 
 \mathbb{R}^{T'}$, $M^{\text{pi}}\in 
 \mathbb{R}^{T'}$, $M^{\text{du}}\in 
 \mathbb{R}^{T'}$ represents information about phoneme, pitch, and duration over the sequence of the same length $T'$. We first introduce the SVS data tokenization, then formulate the SVS fine-tuning on SLM, and lastly demonstrate the conditional flow refinement pipeline.

\vspace{-4mm}

\subsection{SVS Data Tokenization}
\begin{wrapfigure}{r}{0.65\linewidth}
\vspace{-5mm}
    \setlength{\intextsep}{0pt} 
    \setlength{\columnsep}{0pt} 
    \centering
    \includegraphics[width=8.89cm]{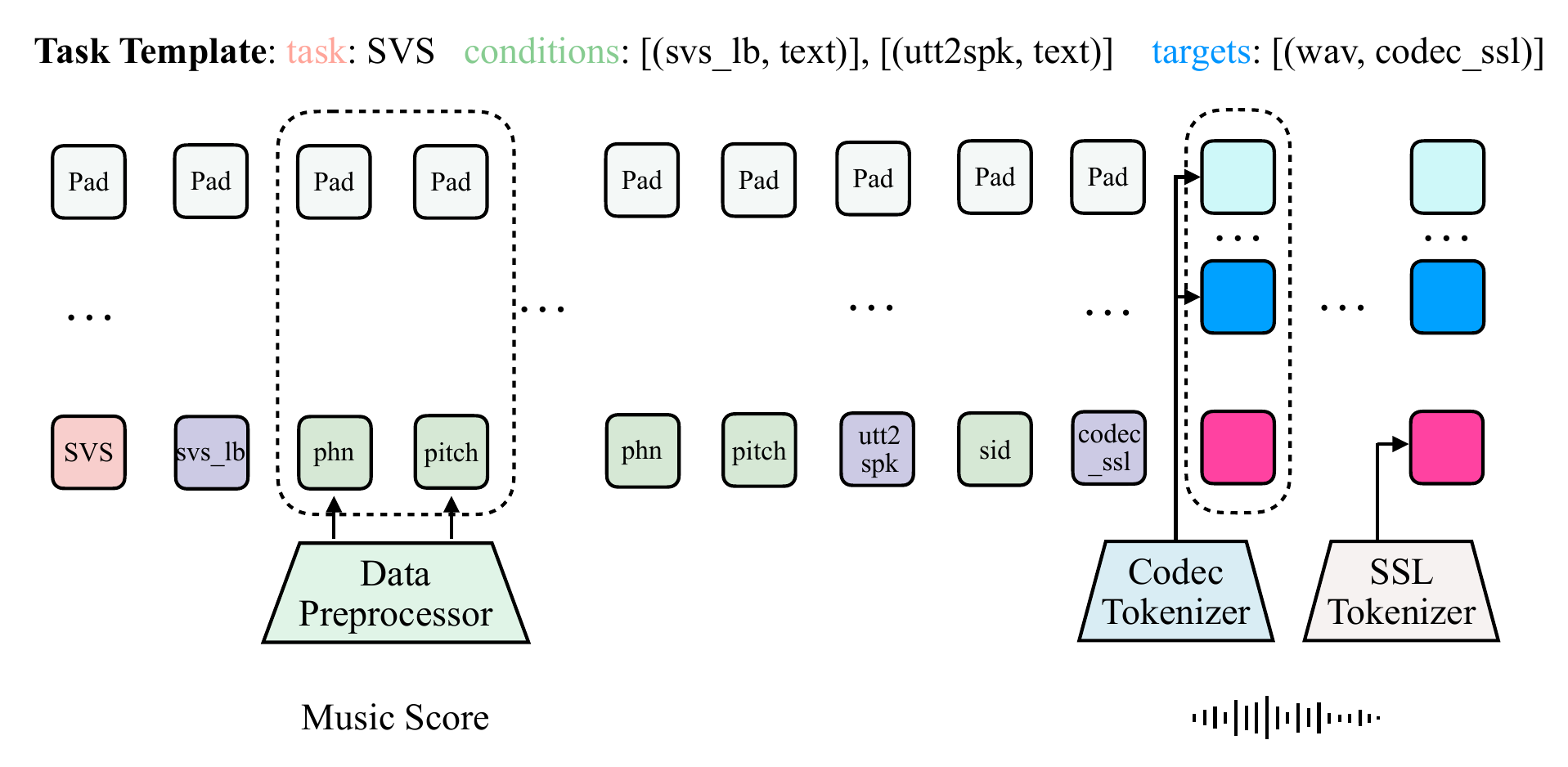}
    \caption{\small SVS fine-tuning task template. The audio codec tokens and SSL tokens are concatenated along the RVQ stream axis.}
    \vspace{2mm}
    \label{fig:svsfinetune}
\end{wrapfigure}
Our pipeline is built upon Espnet-SpeechLM~\cite{tian2025espnetspeechlm, opuslm2025tian}. For SVS, we introduce a new modality called $\text{svs\_lb}$, which consists of frame-level pitch, duration, and phoneme conditions. Each unit in $\text{svs\_lb}$ is a two-element tuple, including a phoneme token and a pitch token with $\text{svs}$ prefix. We use the repeat time of (phn, pitch) tuple to implicitly represent duration, which is calculated as:
$\text{repeat}=(m_{\text{ed}} - m_{\text{st}}) \times \text{fps}$, where $m_\text{ed}$ and $m_\text{st}$ represent the end and start time of an element in $M^{\text{du}}$. This process aligns the annotation with the audio codec sample rate. We show our SVS data tokenization pipeline in Figure~\ref{fig:tokenization}. Our prompt includes the $\text{svs\_lb}$ and $\text{spk\_prompt}$; the target tokens are set to be the concatenation of codec and SSL tokens of singing waveforms.
\vspace{-2mm}

\subsection{Language Model Formulation}
\vspace{-1mm}
For audio representation, we use two types of tokens: the high-level semantics tokens obtained from SSL, and the low-level acoustic tokens from the audio codec. We follow the pre-trained TTS model to use a multi-stream discrete audio representation $\mathbf{s}_f \in \mathbb{N}^{F \times n_q}$ for each frame, where $n_q$ stands for the number of streams, $F$ is the total frame number. Specifically, we concatenate the audio codec tokens and the SSL tokens along the stream dimension, intending to balance their advantages for prediction and acoustic reconstruction. Then, we only keep the codec tokens for decoding. Task and template definitions are shown in Figure~\ref{fig:svsfinetune}.

In SLM, tasks can be uniformly formulated as predicting the target sequence based on the input condition sequence. Adapting this template to SVS, we have the inputs $\mathbf{m} = [\mathbf{m_1}, \mathbf{m_2},...\mathbf{m_F}]$, which is the frame-level music score unit, speaker prompt $\mathbf{p} = [\mathbf{p_1}, \mathbf{p_2},...\mathbf{p_F}]$; and the target frame-level singing feature $\mathbf{s} = [\mathbf{s_1}, \mathbf{s_2},...,\mathbf{s_F}]$. As a token classification task, the fine-tuning objective is to maximize the posterior likelihood $P(\mathbf{s}|\mathbf{m}, \mathbf{p})$ using cross-entropy loss.

\begin{figure*}[t]
\centering
\vspace{-9mm}
\includegraphics[width=13.9cm]{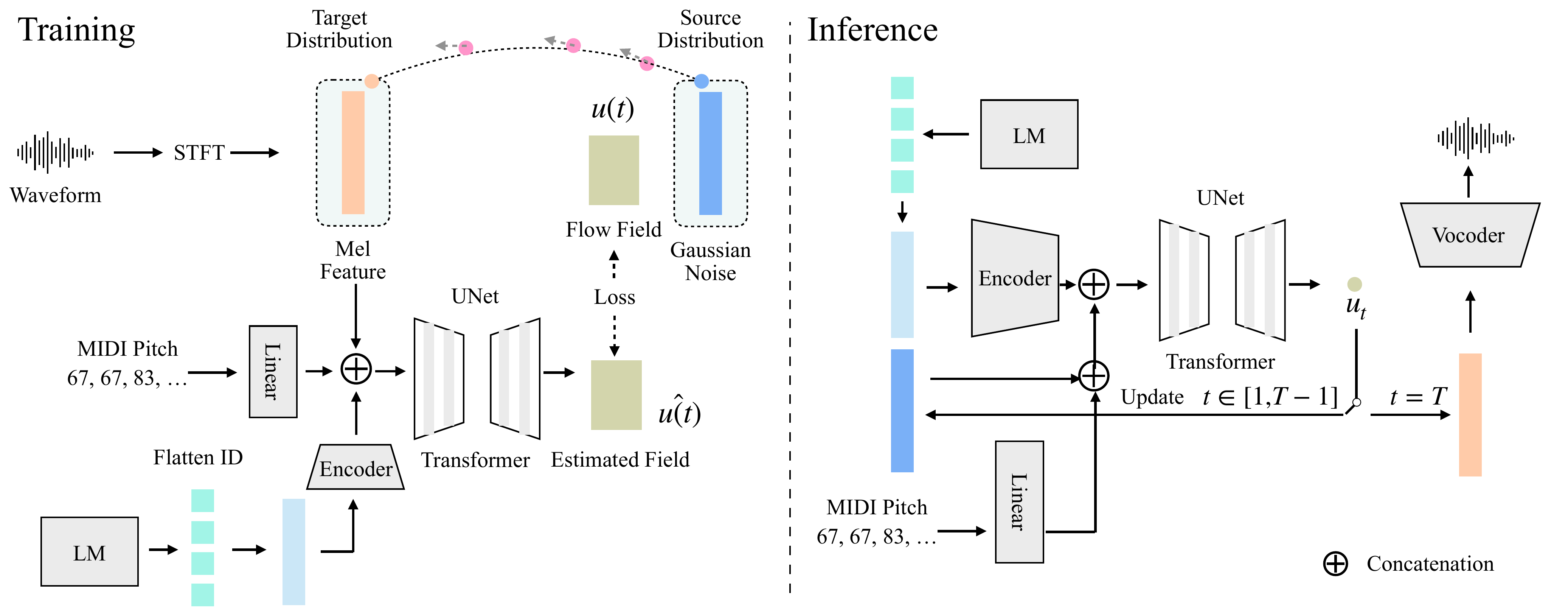}
\vspace{-4mm}
\caption{\small Training and inference process of flow matching.}
\label{fig:flow}
\vspace{-6mm}
\end{figure*}

\subsection{Flow-Based Refinement}
Flow models are a class of generative models that learn a velocity field capable of transporting samples from a source distribution to a target distribution. In this work, we adopt flow matching to train the flow model efficiently and scalably. During training, the model regresses the velocity field along a probabilistic interpolation path by sampling intermediate timesteps $\mathbf{t} \in [0,1]$. At inference, a sample from the source distribution is provided, and the trained velocity field is used to transport it towards the corresponding sample in the target distribution.

Let the source distribution be denoted as $X_0 \sim p$, where $p$ is a standard Gaussian distribution. The corresponding target samples are drawn from $X_1 \sim q$, where $q$ is the distribution of mel features obtained from the target waveforms from STFT. The goal of the flow model $\Psi$ is to learn a smooth mapping from $X_0$ to $X_1$ through a continuous-time velocity field, which follows a continuous-time Markov process. The evolution of a sample over time is governed by:
\begin{equation}
X_{t+h} = \psi_{t+h|t}(X_t), \quad t \in [0,1],
\end{equation}
where $\psi_{t+h|t}$ denotes the transition function over a small time increment $h$. The instantaneous velocity of a point along its trajectory is defined as:
\begin{equation}
\frac{d}{dt} \psi_t(x) = u_t(\psi_t(x)).
\end{equation}
where $u_t$ is the velocity field at time $t$. For flow matching, we adopt a linear interpolation path between source and target samples. This formulation corresponds to the optimal transport path under a kinetic energy minimization constraint,
\begin{equation}
\psi_t(x \mid x_1) = (1 - t)x + t x_1.
\end{equation}
We fuse the LM-predicted codec and pitch signal as additional conditions to make the flow controllable. Let $C$ denote the conditioning input, which contains $\mathbf{s}$ and other optional choices. Let $\theta$ represent the parameters of the conditional flow model. The training objective of Conditional Flow Matching (CFM) is to minimize the squared error between the ground-truth velocity and the model-predicted velocity at a set of points sampled from the path from source distribution to target distribution:
\begin{equation}
\mathcal{L_{\text{CFM}(\theta)}} = \mathbb{E}_{t,(X_0,X_1,C) \sim \pi_{0,1,C}} | u_t(X_t \mid X_1) - u_t^\theta(X_t \mid C) |^2,
\end{equation}
\begin{equation}
u_t^\theta(x \mid c): [0,1] \times \mathbb{R}^d \times \mathbb{R}^k \rightarrow \mathbb{R}^d.
\end{equation}
where $u_t^\theta(x \mid c)$ is the predicted velocity field function. During inference, we solve the corresponding ordinary differential equation (ODE) using a numerical ODE solver, which starts from $X_0 \sim \mathcal{N}(0, I)$, and integrates the velocity field forward in time to obtain the target sample in $X_1$. 
\begin{equation}
\frac{d}{dt} X_t = u_t^\theta(X_t),
\end{equation}
\begin{equation}
Y = \text{Voc}(x_1), \quad x_1 \in X_1.
\end{equation}
The ultimate sample $x_1 \in X_1$ is further converted to a waveform $Y$ using a vocoder,

\begin{table}[t]
\centering
\vspace{-14mm}
\begin{minipage}[t]{0.44\textwidth}
\centering
\caption{\small Comparison of discrete SVS systems.}
\normalsize
\setlength{\tabcolsep}{0.5mm}
\resizebox{\textwidth}{!}{
\begin{tabular}{lcccccc}
\toprule
\multirow{2}{*}{Strategies} & \multicolumn{6}{c}{ACE-Opencpop} \\ 
\cmidrule(r){2-7} 
& F0\_RMSE$\downarrow$ & F0\_CORR$\uparrow$ & MCD$\downarrow$ & PER$\downarrow$ & SingMOS$\uparrow$ & Sheet-SSQA$\uparrow$  \\ 
\midrule \midrule
XiaoiceSing & 71.67 & 0.62 & 11.47 & 0.09 & 3.88 & 3.62 \\
TokSing & 55.83 & 0.67 & 6.77 & 0.19 & 4.08 & 3.89 \\
LM + Flow1 + Voc & 62.79 & 0.60 & 7.86 & 0.36 & 4.09 & 3.79 \\
\bottomrule
\end{tabular}}
\label{tab_compare}
\end{minipage}
\hfill
\begin{minipage}[t]{0.55\textwidth}
\centering
\caption{\small Ablation on recipe designs.}
\small
\setlength{\tabcolsep}{0.5mm}
\resizebox{\textwidth}{!}{
\begin{tabular}{lccccccc}
\toprule
\multirow{2}{*}{Conditions} & \multicolumn{7}{c}{ACE-Opencpop} \\ 
\cmidrule(r){2-8} 
& F0\_RMSE$\downarrow$ & F0\_CORR$\uparrow$ & MCD$\downarrow$ & PER$\downarrow$ & Singer-Sim$\uparrow$ & SingMOS$\uparrow$ & Sheet-SSQA$\uparrow$ \\ 
\midrule \midrule
CD Resynthesis & 51.38 & 0.73 & 5.84 & 0.19 & 0.67 & 3.95 & 3.78 \\
LM + CD  & 62.90 & 0.60 & 8.26 & 0.56 & 0.49 & 3.65 & 3.08 \\
LM + Flow1 + CD & 61.51 & 0.60 & 8.44 & 0.45 & 0.49 & 3.64 & 3.08\\
LM + Flow1 + Voc & 62.79 & 0.60 & 7.86 & 0.36 & 0.61 & 4.09 & 3.79 \\
LM + Flow2 + Voc  & 62.52 & 0.62 & 7.66 & 0.42 & 0.56 & 3.95 & 3.55 \\
\bottomrule
\end{tabular}}
\label{tab_abla}
\end{minipage}
\vspace{-3mm}
\end{table}

\vspace{-4mm}
\section{Experiment}
\vspace{-2mm}
\subsection{Corpus and Parameters Setups}
\noindent \textbf{ACE-Opencpop} is a synthetic corpus that inherits the song list from 5.2-hour Mandarin female singing corpus Opencpop, but curates the singing of 30 additional singers using ACE Studio with manual tuning, resulting in the largest, 135-hour opensourced SVS corpus. Detailed parameters setup for SLM fine-tuning and flow matching are shown in appendix~\ref{sec_parasetup}.

\vspace{-2mm}
\subsection{Results Analysis and Ablations}
\vspace{-1mm}
\label{sec_res}
\noindent\textbf{Explanation of the abbreviations used in Tab.~\ref{tab_compare} and Tab.~\ref{tab_abla}.} \textbf{XiaoiceSing} uses music score information to predict discrete tokens and train a vocoder on discrete tokens to waveform. \textbf{TokSing} also builds on a discrete NAR architecture, but further introduces a melody predictor and a music enhancer to improve pitch precision. \textbf{CD Resynthesis} denotes directly encoding and decoding singing waveforms using a speech-pretrained codec model. \textbf{Flow1} refers to flow matching conditioned on the LM-predicted codec features, while \textbf{Flow2} conditions on both the LM-predicted codec features and the pitch tensor. \textbf{+CD} indicates that the flow output is in the codec embedding space and is converted to waveforms via a pretrained codec decoder. \textbf{+Voc} indicates that the flow output is in the mel-spectrogram space and is converted to waveforms via a vocoder.

\noindent\textbf{Metrics used in Tab.~\ref{tab_compare} and Tab.~\ref{tab_abla}.} \textbf{F0\_RMSE} and \textbf{F0\_CORR}~\cite{hayashi2020espnet} measure the root mean squared error and correlation between the fundamental frequency (F0) of the synthesized and reference singing signals, focusing on pitch accuracy. \textbf{MCD}~\cite{kubichek1993mel} quantifies the spectral distance between the generated and ground-truth audio using mel-cepstral coefficients. \textbf{PER} denotes phoneme error rate. \textbf{SingMOS}~\cite{singmos} and \textbf{Sheet-SSQA}~\cite{huang2024mos} is a pseudo MOS predictor based on a 5-point mean opinion score scale.

\noindent\textbf{Quantitative Analysis.}
As shown in Tab.~\ref{tab_compare}, our SLM-based SVS pipeline achieves performance comparable to state-of-the-art discrete SVS models. Compare with XiaoiceSing~\cite{lu2020xiaoicesing}, our model performs better in pitch-related f0 metrics and overall singing quality, as indicated by pseudo MOS. While the pitch accuracy (reflected by the f0-related metrics) slightly lags behind TokSing~\cite{wu2024toksing}, the overall singing quality matches or even surpasses it, demonstrating the effectiveness of our architectural design and the strong downstream generalization capability of SLM. The ablation study highlights the impact of using mel versus codec features as the flow output space: mel features appear easier to model, as indicated by the greater improvement over the LM+CD baseline. Also, as the speech-pretrained codec model leads to information loss in resynthesis already, it sets an upper bound for using the codec feature to form the flow output space. Moreover, incorporating pitch information into flow matching yields a modest gain in pitch fidelity, which is shown by the comparison between LM + Flow1 + Voc and LM + Flow2 + Voc.

\vspace{-4mm}
\section{Conclusion}
\vspace{-2mm}
In this paper, we present a pipeline that adapts a TTS-pretrained SLM for the SVS task, achieving performance on par with state-of-the-art discrete SVS methods. This demonstrates the strong generalizability of SLMs in low-resource downstream settings and points to promising directions for future multi-task SLM research.

\noindent\textbf{Acknowledgement.} Experiments of this work used the Bridges2 at PSC and Delta/DeltaAI NCSA computing systems through allocation CIS210014 from the ACCESS program, supported by NSF grants 2138259, 2138286, 2138307, 2137603, and 2138296.

{
\small
\bibliographystyle{plainnat}
\bibliography{custom}

@article{tian2025espnetspeechlm,
  author       = {Jinchuan Tian and
                  Jiatong Shi and
                  William Chen and
                  Siddhant Arora and
                  Yoshiki Masuyama and
                  Takashi Maekaku and
                  Yihan Wu and
                  Junyi Peng and
                  Shikhar Bharadwaj and
                  Yiwen Zhao and
                  Samuele Cornell and
                  Yifan Peng and
                  Xiang Yue and
                  Chao{-}Han Huck Yang and
                  Graham Neubig and
                  Shinji Watanabe},
  title        = {ESPnet-SpeechLM: An Open Speech Language Model Toolkit},
  journal      = {CoRR},
  volume       = {abs/2502.15218},
  year         = {2025},
  url          = {https://doi.org/10.48550/arXiv.2502.15218},
}

@inproceedings{saino2006hmm,
  title={An HMM-based singing voice synthesis system.},
  author={Saino, Keijiro and Zen, Heiga and Nankaku, Yoshihiko and Lee, Akinobu and Tokuda, Keiichi},
  booktitle={INTERSPEECH},
  pages={2274--2277},
  year={2006}
}

@article{lu2020xiaoicesing,
  title={Xiaoicesing: A high-quality and integrated singing voice synthesis system},
  author={Lu, Peiling and Wu, Jie and Luan, Jian and Tan, Xu and Zhou, Li},
  journal={arXiv preprint arXiv:2006.06261},
  year={2020}
}

@article{chu2024qwen2,
  title={Qwen2-audio technical report},
  author={Chu, Yunfei and Xu, Jin and Yang, Qian and Wei, Haojie and Wei, Xipin and Guo, Zhifang and Leng, Yichong and Lv, Yuanjun and He, Jinzheng and Lin, Junyang and others},
  journal={arXiv preprint arXiv:2407.10759},
  year={2024}
}

@article{abouelenin2025phi,
  title={Phi-4-mini technical report: Compact yet powerful multimodal language models via mixture-of-loras},
  author={Abouelenin, Abdelrahman and Ashfaq, Atabak and Atkinson, Adam and Awadalla, Hany and Bach, Nguyen and Bao, Jianmin and Benhaim, Alon and Cai, Martin and Chaudhary, Vishrav and Chen, Congcong and others},
  journal={arXiv preprint arXiv:2503.01743},
  year={2025}
}

@article{ding2025kimi,
  title={Kimi-audio technical report},
  author={Ding, Ding and Ju, Zeqian and Leng, Yichong and Liu, Songxiang and Liu, Tong and Shang, Zeyu and Shen, Kai and Song, Wei and Tan, Xu and Tang, Heyi and others},
  journal={arXiv preprint arXiv:2504.18425},
  year={2025}
}

@article{wang2023neural,
  title={Neural codec language models are zero-shot text to speech synthesizers},
  author={Wang, Chengyi and Chen, Sanyuan and Wu, Yu and Zhang, Ziqiang and Zhou, Long and Liu, Shujie and Chen, Zhuo and Liu, Yanqing and Wang, Huaming and Li, Jinyu and others},
  journal={arXiv preprint arXiv:2301.02111},
  year={2023}
}

@misc{kharitonov2023speakreadprompthighfidelity,
      title={Speak, Read and Prompt: High-Fidelity Text-to-Speech with Minimal Supervision}, 
      author={Eugene Kharitonov and Damien Vincent and Zalán Borsos and Raphaël Marinier and Sertan Girgin and Olivier Pietquin and Matt Sharifi and Marco Tagliasacchi and Neil Zeghidour},
      year={2023},
      eprint={2302.03540},
      archivePrefix={arXiv},
      primaryClass={cs.SD},
      url={https://arxiv.org/abs/2302.03540}, 
}

@article{kingma2018glow,
  title={Glow: Generative flow with invertible 1x1 convolutions},
  author={Kingma, Durk P and Dhariwal, Prafulla},
  journal={Advances in neural information processing systems},
  volume={31},
  year={2018}
}

@inproceedings{prenger2019waveglow,
  title={Waveglow: A flow-based generative network for speech synthesis},
  author={Prenger, Ryan and Valle, Rafael and Catanzaro, Bryan},
  booktitle={ICASSP 2019-2019 IEEE International Conference on Acoustics, Speech and Signal Processing (ICASSP)},
  pages={3617--3621},
  year={2019},
  organization={IEEE}
}

@inproceedings{popov2021grad,
  title={Grad-tts: A diffusion probabilistic model for text-to-speech},
  author={Popov, Vadim and Vovk, Ivan and Gogoryan, Vladimir and Sadekova, Tasnima and Kudinov, Mikhail},
  booktitle={International conference on machine learning},
  pages={8599--8608},
  year={2021},
  organization={PMLR}
}

@article{dinh2016density,
  title={Density estimation using real nvp},
  author={Dinh, Laurent and Sohl-Dickstein, Jascha and Bengio, Samy},
  journal={arXiv preprint arXiv:1605.08803},
  year={2016}
}

@inproceedings{strauss2023improved,
  title={Improved normalizing flow-based speech enhancement using an all-pole gammatone filterbank for conditional input representation},
  author={Strauss, Martin and Torcoli, Matteo and Edler, Bernd},
  booktitle={2022 IEEE Spoken Language Technology Workshop (SLT)},
  pages={444--450},
  year={2023},
  organization={IEEE}
}

@article{tong2023conditional,
  title={Conditional flow matching: Simulation-free dynamic optimal transport},
  author={Tong, Alexander and Malkin, Nikolay and Huguet, Guillaume and Zhang, Yanlei and Rector-Brooks, Jarrid and Fatras, Kilian and Wolf, Guy and Bengio, Yoshua},
  journal={arXiv preprint arXiv:2302.00482},
  volume={2},
  number={3},
  year={2023}
}

@misc{zhang2025inspiremusicintegratingsuperresolution,
      title={InspireMusic: Integrating Super Resolution and Large Language Model for High-Fidelity Long-Form Music Generation}, 
      author={Chong Zhang and Yukun Ma and Qian Chen and Wen Wang and Shengkui Zhao and Zexu Pan and Hao Wang and Chongjia Ni and Trung Hieu Nguyen and Kun Zhou and Yidi Jiang and Chaohong Tan and Zhifu Gao and Zhihao Du and Bin Ma},
      year={2025},
      eprint={2503.00084},
      archivePrefix={arXiv},
      primaryClass={cs.SD},
      url={https://arxiv.org/abs/2503.00084}, 
}

@misc{bai2024seedmusicunifiedframeworkhigh,
      title={Seed-Music: A Unified Framework for High Quality and Controlled Music Generation}, 
      author={Ye Bai and Haonan Chen and Jitong Chen and Zhuo Chen and Yi Deng and Xiaohong Dong and Lamtharn Hantrakul and Weituo Hao and Qingqing Huang and Zhongyi Huang and Dongya Jia and Feihu La and Duc Le and Bochen Li and Chumin Li and Hui Li and Xingxing Li and Shouda Liu and Wei-Tsung Lu and Yiqing Lu and Andrew Shaw and Janne Spijkervet and Yakun Sun and Bo Wang and Ju-Chiang Wang and Yuping Wang and Yuxuan Wang and Ling Xu and Yifeng Yang and Chao Yao and Shuo Zhang and Yang Zhang and Yilin Zhang and Hang Zhao and Ziyi Zhao and Dejian Zhong and Shicen Zhou and Pei Zou},
      year={2024},
      eprint={2409.09214},
      archivePrefix={arXiv},
      primaryClass={cs.SD},
      url={https://arxiv.org/abs/2409.09214}, 
}

@inproceedings{visinger2,
  author       = {Yongmao Zhang and
                  Heyang Xue and
                  Hanzhao Li and
                  Lei Xie and
                  Tingwei Guo and
                  Ruixiong Zhang and
                  Caixia Gong},
  editor       = {Naomi Harte and
                  Julie Carson{-}Berndsen and
                  Gareth Jones},
  title        = {VISinger2: High-Fidelity End-to-End Singing Voice Synthesis Enhanced
                  by Digital Signal Processing Synthesizer},
  booktitle    = {24th Annual Conference of the International Speech Communication Association,
                  Interspeech 2023, Dublin, Ireland, August 20-24, 2023},
  pages        = {4444--4448},
  publisher    = {{ISCA}},
  year         = {2023},
  url          = {https://doi.org/10.21437/Interspeech.2023-391},
}

@inproceedings{wu2024toksing,
  author       = {Yuning Wu and
                  Chunlei Zhang and
                  Jiatong Shi and
                  Yuxun Tang and
                  Shan Yang and
                  Qin Jin},
  editor       = {Itshak Lapidot and
                  Sharon Gannot},
  title        = {TokSing: Singing Voice Synthesis based on Discrete Tokens},
  booktitle    = {25th Annual Conference of the International Speech Communication Association,
                  Interspeech 2024, Kos, Greece, September 1-5, 2024},
  publisher    = {{ISCA}},
  year         = {2024},
}

@article{opuslm2025tian,
  author       = {Jinchuan Tian and
                  William Chen and
                  Yifan Peng and
                  Jiatong Shi and
                  Siddhant Arora and
                  Shikhar Bharadwaj and
                  Takashi Maekaku and
                  Yusuke Shinohara and
                  Keita Goto and
                  Xiang Yue and
                  Huck Yang and
                  Shinji Watanabe},
  title        = {OpusLM: {A} Family of Open Unified Speech Language Models},
  journal      = {CoRR},
  volume       = {abs/2506.17611},
  year         = {2025},
}

@inproceedings{espnet-codec,
  author       = {Jiatong Shi and
                  Jinchuan Tian and
                  Yihan Wu and
                  Jee{-}Weon Jung and
                  Jia Qi Yip and
                  Yoshiki Masuyama and
                  William Chen and
                  Yuning Wu and
                  Yuxun Tang and
                  Massa Baali and
                  Dareen Alharthi and
                  Dong Zhang and
                  Ruifan Deng and
                  Tejes Srivastava and
                  Haibin Wu and
                  Alexander H. Liu and
                  Bhiksha Raj and
                  Qin Jin and
                  Ruihua Song and
                  Shinji Watanabe},
  title        = {ESPnet-Codec: Comprehensive Training and Evaluation of Neural Codecs
                  For Audio, Music, and Speech},
  booktitle    = {{IEEE} Spoken Language Technology Workshop, {SLT} 2024, Macao, December
                  2-5, 2024},
  pages        = {562--569},
  publisher    = {{IEEE}},
  year         = {2024},
}

@inproceedings{hifigan,
  author       = {Jungil Kong and
                  Jaehyeon Kim and
                  Jaekyoung Bae},
  editor       = {Hugo Larochelle and
                  Marc'Aurelio Ranzato and
                  Raia Hadsell and
                  Maria{-}Florina Balcan and
                  Hsuan{-}Tien Lin},
  title        = {HiFi-GAN: Generative Adversarial Networks for Efficient and High Fidelity
                  Speech Synthesis},
  booktitle    = {Advances in Neural Information Processing Systems 33: Annual Conference
                  on Neural Information Processing Systems 2020, NeurIPS 2020, December
                  6-12, 2020, virtual},
  year         = {2020},
}

@inproceedings{promptsinger,
  author       = {Yongqi Wang and
                  Ruofan Hu and
                  Rongjie Huang and
                  Zhiqing Hong and
                  Ruiqi Li and
                  Wenrui Liu and
                  Fuming You and
                  Tao Jin and
                  Zhou Zhao},
  editor       = {Kevin Duh and
                  Helena G{\'{o}}mez{-}Adorno and
                  Steven Bethard},
  title        = {Prompt-Singer: Controllable Singing-Voice-Synthesis with Natural Language
                  Prompt},
  booktitle    = {Proceedings of the 2024 Conference of the North American Chapter of
                  the Association for Computational Linguistics: Human Language Technologies
                  (Volume 1: Long Papers), {NAACL} 2024, Mexico City, Mexico, June 16-21,
                  2024},
  pages        = {4780--4794},
  publisher    = {Association for Computational Linguistics},
  year         = {2024},
}

@article{TTSong,
  author       = {Ruiqi Li and
                  Zhiqing Hong and
                  Yongqi Wang and
                  Lichao Zhang and
                  Rongjie Huang and
                  Siqi Zheng and
                  Zhou Zhao},
  title        = {Accompanied Singing Voice Synthesis with Fully Text-controlled Melody},
  journal      = {CoRR},
  volume       = {abs/2407.02049},
  year         = {2024},
  url          = {https://doi.org/10.48550/arXiv.2407.02049},
}

@article{LLFMvoice,
	doi = {10.20944/preprints202504.1831.v1},
	url = {https://doi.org/10.20944/preprints202504.1831.v1},
	year = 2025,
	month = {April},
	publisher = {Preprints},
	author = {Yanze Wang and Xuming Han and Shuai Lv and Ting Zhou and Yali Chu},
	title = {LLFM-Voice: Emotionally Expressive Speech and Singing Voice Synthesis with Large Language Models via Flow Matching},
	journal = {Preprints}
}

@inproceedings{hayashi2020espnet,
  title={ESPnet-{TTS}: Unified, reproducible, and integratable open source end-to-end text-to-speech toolkit},
  author={Hayashi, Tomoki and Yamamoto, Ryuichi and Inoue, Katsuki and Yoshimura, Takenori and Watanabe, Shinji and Toda, Tomoki and Takeda, Kazuya and Zhang, Yu and Tan, Xu},
  booktitle={ICASSP 2020-2020 IEEE international conference on acoustics, speech and signal processing (ICASSP)},
  pages={7654--7658},
  year={2020},
  organization={IEEE}
}

@inproceedings{kubichek1993mel,
  title={Mel-cepstral distance measure for objective speech quality assessment},
  author={Kubichek, Robert},
  booktitle={Proceedings of IEEE pacific rim conference on communications computers and signal processing},
  volume={1},
  pages={125--128},
  year={1993},
  organization={IEEE}
}

@article{huang2024mos,
  title={Mos-bench: Benchmarking generalization abilities of subjective speech quality assessment models},
  author={Huang, Wen-Chin and Cooper, Erica and Toda, Tomoki},
  journal={arXiv preprint arXiv:2411.03715},
  year={2024}
}

@article{singmos,
  author       = {Yuxun Tang and
                  Jiatong Shi and
                  Yuning Wu and
                  Qin Jin},
  title        = {SingMOS: An extensive Open-Source Singing Voice Dataset for {MOS}
                  Prediction},
  journal      = {CoRR},
  volume       = {abs/2406.10911},
  year         = {2024},
  url          = {https://doi.org/10.48550/arXiv.2406.10911}
}
}


\appendix

\section{Related Works}
\label{sec_rw}
\subsection{Speech Language Models}

Large language models have witnessed significant advancements in recent years, primarily driven by increased model scale, enhanced data availability, and improved training paradigms. Their ability to learn unified representations across modalities has enabled progress on a wide range of tasks. 

Recent works have extended this paradigm to the audio domain, resulting in speech language models (SLMs) that treat speech as a sequence of discrete tokens, analogous to natural language~\cite{abouelenin2025phi, chu2024qwen2, ding2025kimi, wang2023neural, kharitonov2023speakreadprompthighfidelity}. SLMs demonstrate strong transferability from general pretraining to specific downstream tasks, 
 using techniques like fine-tuning~\cite{ding2025kimi,chu2024qwen2} and LORAs~\cite{abouelenin2025phi}, while pretraining is typically conducted on large-scale, general-purpose corpora, domain adaptation is feasible through fine-tuning on curated task-specific datasets.

Previous works also explore using a language model to generate music~\cite {zhang2025inspiremusicintegratingsuperresolution, bai2024seedmusicunifiedframeworkhigh}. However, training the language model on singing requires a large amount of data, which is always closed-source and hard to access by the community. Our work adapt TTS-pretrained SLM to a low-resource SVS setting.

\subsection{Singing Voice Synthesis}

Singing voice synthesis (SVS) aims to generate expressive and intelligible singing voices from structured musical inputs, typically including phoneme-level lyrics, MIDIs, and note durations. Compared to TTS, SVS requires a higher degree of temporal precision and pitch accuracy to maintain musicality. This renders SVS more sensitive to the alignment between input conditions and the generated acoustic output. Early approaches in SVS were based on concatenative and HMM-based methods~\cite{saino2006hmm}, while recent work has shifted towards neural vocoder-based systems~\cite{lu2020xiaoicesing}, including encoder-decoder architectures and end-to-end frameworks~\cite{visinger2}.

Recent studies have begun leveraging large language models (LLMs) for singing voice synthesis (SVS) and text-to-song generation, enabling more flexible and controllable singing beyond traditional alignment-based methods. Prompt-Singer~\cite{promptsinger} introduces a prompt-based SVS system that controls vocal style (e.g., timbre, range) via natural language, using a decoder-only transformer with a range–melody decoupled pitch representation for intuitive style manipulation. MelodyLM / TTSong~\cite{TTSong} advances toward fully text-driven song generation by predicting melody representations from lyrics and textual descriptions, integrating LLM-based melody modeling with diffusion-based accompaniment synthesis. LLFM-Voice~\cite{LLFMvoice} unifies expressive speech and singing synthesis using an LLM front-end and flow-matching acoustic model, achieving smoother emotional expression and higher-fidelity vocal rendering.

Our work differs by leveraging a general-purpose speech language model, pre-trained on TTS data, and adapting it to SVS through token-level modeling and conditional refinement, enabling the complex SVS to be a subtask of a unified model.

\subsection{Flow-Based Models}
Flow-based generative models, such as RealNVP~\cite{dinh2016density} and Glow~\cite{kingma2018glow}, are a class of invertible neural networks that learn data distributions through a sequence of bijective transformations. These classical flow models enable exact likelihood estimation and efficient sampling, and have been successfully applied to high-fidelity speech and audio generation tasks, including vocoding~\cite{prenger2019waveglow}, speech enhancement~\cite{strauss2023improved}, and expressive speech synthesis~\cite{popov2021grad}.

In contrast, flow-matching approaches~\cite{tong2023conditional} are ODE-based methods conceptually closer to diffusion models (which are SDE-based). Instead of performing explicit density estimation like classical flows, flow-matching trains conditional flows via velocity field learning, providing a scalable and flexible framework for conditional generation.

In this work, we adopt a conditional flow-matching model to generate mel-spectrogram features from Gaussian noise, conditioned on LM-predicted codec embeddings as well as musical pitch labels. This approach refines the acoustic realism and pitch fidelity of generated singing voices while avoiding the computational overhead of traditional flow-based likelihood estimation.



\section{Model Parameter Setups}
\label{sec_parasetup}
\subsection{Language Model Finetuning.}
We finetune the model using \texttt{DeepSpeed} with mixed-precision (\texttt{FP16}) training, Adam optimizer (\(\beta_1=0.9,\ \beta_2=0.95\), learning rate \(5\times10^{-6}\)), and ZeRO stage-2 optimization for memory efficiency. A Warmup--Cosine learning rate scheduler with \(7\times10^{5}\) total steps (minimum LR ratio \(0.3\)) is employed. Gradient clipping is set to \(1.0\). Contiguous memory and communication-overlap optimizations are enabled to ensure stable and scalable training.

\subsection{Flow Matching.} We train the model for a total of \(30\) epochs with a batch size of \(64\). The learning rate is scheduled to decay linearly from \(3 \times 10^{-4}\) to \(1 \times 10^{-4}\) between steps \(2\times10^{5}\) and \(5\times10^{5}\).

\end{document}